\newcommand{\DoPrePrint}{0} 

\ifnum\DoPrePrint=1
  \RequirePackage{lineno} 
  \documentclass[aps,prl,preprint,showpacs,superscriptaddress,nofootinbib,linenumbers]{revtex4}

  \usepackage{lineno}
\else
  \documentclass[aps,prl,twocolumn,showpacs,superscriptaddress,nofootinbib]{revtex4}
\fi

\usepackage{amsmath}   
\usepackage{graphicx}  
\usepackage{xspace}
\usepackage{units}

\usepackage{hyperref}
\usepackage{multirow}

\newcommand{\minerva}{MINERvA\xspace}
\newcommand{\minos}{MINOS\xspace}

\newcommand{\numi}{NuMI\xspace}
\newcommand{\genie}{GENIE\xspace}
\newcommand{\nuwro}{NuWro\xspace}
\newcommand{\geant}{GEANT4\xspace}

\newcommand{\numu}{\ensuremath{\nu_{\mu}}\xspace}
\newcommand{\numubar}{\ensuremath{\bar{\nu}_{\mu}}\xspace}
\newcommand{\dsdq}{\ensuremath{\mathrm{d}\sigma/\mathrm{d}Q^2}\xspace}
\newcommand{\QSq}{\ensuremath{Q^{2}}\xspace}
\newcommand{\QSqproton}{\ensuremath{Q^{2}_{QE,p}}\xspace}
\newcommand{\QSqmuon}{\ensuremath{Q^{2}_{QE,\mu}}\xspace}

\newcommand{\xsecproton}{\ensuremath{\mathrm{d}\sigma/\mathrm{d}Q^{2}_{QE,p}}\xspace}
\newcommand{\enu}{\ensuremath{E_{\nu}}\xspace}

\newcommand{\sizecheck}{0} 
\newcommand{\PRLsupp}{1}   

\ifnum\PRLsupp=0
  
\else
  
\fi

\newif\ifpdf
\ifx\pdfoutput\undefined
   \pdffalse
\else
   \pdfoutput=1
   \pdftrue
\fi
\ifpdf
   \usepackage{graphicx}
   \usepackage{epstopdf}
\else
   \usepackage{graphicx}
\fi

\begin{document}

\ifnum\DoPrePrint=1
\linenumbers
\fi

\title{Measurement of muon plus proton final states in \numu Interactions on Hydrocarbon at $\langle\enu\rangle$~=~4.2~GeV}



\newcommand{\deceased}{Deceased}

\newcommand{\wroclaw}{Institute of Theoretical Physics, Wroc\l aw University, Wroc\l aw, Poland}    


\newcommand{\Rutgers}{Rutgers, The State University of New Jersey, Piscataway, New Jersey 08854, USA}
\newcommand{\Hampton}{Hampton University, Dept. of Physics, Hampton, VA 23668, USA}
\newcommand{\Dortmund}{Institute of Physics, Dortmund University, 44221, Germany }
\newcommand{\Otterbein}{Department of Physics, Otterbein University, 1 South Grove Street, Westerville, OH, 43081 USA}
\newcommand{\JMU}{James Madison University, Harrisonburg, Virginia 22807, USA}
\newcommand{\Florida}{University of Florida, Department of Physics, Gainesville, FL 32611}
\newcommand{\UCIrvine}{Department of Physics and Astronomy, University of California, Irvine, Irvine, California 92697-4575, USA}
\newcommand{\CBPF}{Centro Brasileiro de Pesquisas F\'{i}sicas, Rua Dr. Xavier Sigaud 150, Urca, Rio de Janeiro, RJ, 22290-180, Brazil}
\newcommand{\PUCP}{Secci\'{o}n F\'{i}sica, Departamento de Ciencias, Pontificia Universidad Cat\'{o}lica del Per\'{u}, Apartado 1761, Lima, Per\'{u}}
\newcommand{\INRM}{Institute for Nuclear Research of the Russian Academy of Sciences, 117312 Moscow, Russia}
\newcommand{\Jlab}{Jefferson Lab, 12000 Jefferson Avenue, Newport News, VA 23606, USA}
\newcommand{\Pittsburgh}{Department of Physics and Astronomy, University of Pittsburgh, Pittsburgh, Pennsylvania 15260, USA}
\newcommand{\Guanajuato}{Campus Le\'{o}n y Campus Guanajuato, Universidad de Guanajuato, Lascurain de Retana No. 5, Col. Centro. Guanajuato 36000, Guanajuato M\'{e}xico.}
\newcommand{\Athens}{Department of Physics, University of Athens, GR-15771 Athens, Greece}
\newcommand{\Tufts}{Physics Department, Tufts University, Medford, Massachusetts 02155, USA}
\newcommand{\WM}{Department of Physics, College of William \& Mary, Williamsburg, Virginia 23187, USA}
\newcommand{\FNAL}{Fermi National Accelerator Laboratory, Batavia, Illinois 60510, USA}
\newcommand{\Purdue}{Department of Chemistry and Physics, Purdue University Calumet, Hammond, Indiana 46323, USA}
\newcommand{\MCLA}{Massachusetts College of Liberal Arts, 375 Church Street, North Adams, MA 01247}
\newcommand{\UMD}{Department of Physics, University of Minnesota -- Duluth, Duluth, Minnesota 55812, USA}
\newcommand{\Northwestern}{Northwestern University, Evanston, Illinois 60208}
\newcommand{\UNI}{Universidad Nacional de Ingenier\'{i}a, Apartado 31139, Lima, Per\'{u}}
\newcommand{\Rochester}{University of Rochester, Rochester, New York 14610 USA}
\newcommand{\Austin}{Department of Physics, University of Texas, 1 University Station, Austin, Texas 78712, USA}
\newcommand{\USM}{Departamento de F\'{i}sica, Universidad T\'{e}cnica Federico Santa Mar\'{i}a, Avda. Espa\~{n}a 1680 Casilla 110-V, Valpara\'{i}so, Chile}
\newcommand{\Geneva}{University of Geneva, Geneva, Switzerland}
\newcommand{\Chicago}{Enrico Fermi Institute, University of Chicago, Chicago, IL 60637 USA}
\newcommand{\keppelThanks}{\thanks{now at the Thomas Jefferson National Accelerator Facility, Newport News, VA 23606 USA}}
\newcommand{\giulianoThanks}{\thanks{now at Vrije Universiteit Brussel, Pleinlaan 2, B-1050 Brussels, Belgium}}
\newcommand{\schulteThanks}{\thanks{now at Temple University, Philadelphia, Pennsylvania 19122, USA}}
\newcommand{\jwaldingThanks}{\thanks{now at Dept. Physics, Royal Holloway, University of London, UK}}
\newcommand{\ticeThanks}{now at Argonne National Laboratory, Argonne, IL 60439, USA }
\newcommand{\bmeThanks}{now at SLAC National Accelerator Laboratory, Stanford, California 94309 USA}
\newcommand{\waltonThanks}{now at Fermi National Accelerator Laboratory, Batavia, IL 60510 USA} 
\newcommand{\lazaThanks}{also at Department of Physics, University of Antananarivo, Madagascar}
\newcommand{\janThanks}{also at Institute of Theoretical Physics, Wroc\l aw University, Wroc\l aw, Poland}

\author{T.~Walton}\thanks{\waltonThanks}           \affiliation{\Hampton}
\author{M.~Betancourt}        \affiliation{\FNAL}
\author{L.~Aliaga}                        \affiliation{\WM}  \affiliation{\PUCP}
\author{O.~Altinok}                       \affiliation{\Tufts}
\author{A.~Bodek}                         \affiliation{\Rochester}
\author{A.~Bravar}                        \affiliation{\Geneva}
\author{H.~Budd}                          \affiliation{\Rochester}
\author{M.~J.~Bustamante}                \affiliation{\PUCP}
\author{A.~Butkevich}                     \affiliation{\INRM}
\author{D.A.~Martinez~Caicedo}            \affiliation{\CBPF}  \affiliation{\FNAL}
\author{M.F.~Carneiro}                    \affiliation{\CBPF}
\author{C.M.~Castromonte}                 \affiliation{\CBPF}
\author{M.E.~Christy}                     \affiliation{\Hampton}
\author{J.~Chvojka}                       \affiliation{\Rochester}
\author{H.~da~Motta}                      \affiliation{\CBPF}
\author{M.~Datta}                         \affiliation{\Hampton}
\author{J.~Devan}                         \affiliation{\WM}
\author{S.A.~Dytman}                      \affiliation{\Pittsburgh}
\author{G.A.~D\'{i}az~}                   \affiliation{\PUCP}
\author{B.~Eberly}\thanks{\bmeThanks}   \affiliation{\Pittsburgh}
\author{J.~Felix}                         \affiliation{\Guanajuato}
\author{L.~Fields}                        \affiliation{\Northwestern}
\author{R.~Fine}                          \affiliation{\Rochester}
\author{G.A.~Fiorentini}                  \affiliation{\CBPF}
\author{A.M.~Gago}                        \affiliation{\PUCP}
\author{H.~Gallagher}                     \affiliation{\Tufts}
\author{R.~Gran}                          \affiliation{\UMD}
\author{D.A.~Harris}                      \affiliation{\FNAL}
\author{A.~Higuera}                       \affiliation{\Rochester}  \affiliation{\Guanajuato}
\author{K.~Hurtado}                       \affiliation{\CBPF}  \affiliation{\UNI}
\author{J.~Kleykamp}                      \affiliation{\Rochester}
\author{M.~Kordosky}                      \affiliation{\WM}
\author{S.A.~Kulagin}                     \affiliation{\INRM}
\author{T.~Le}                            \affiliation{\Rutgers}
\author{E.~Maher}                         \affiliation{\MCLA}
\author{S.~Manly}                         \affiliation{\Rochester}
\author{W.A.~Mann}                        \affiliation{\Tufts}
\author{C.M.~Marshall}                    \affiliation{\Rochester}
\author{C.~Martin~Mari}                   \affiliation{\Geneva}
\author{K.S.~McFarland}                   \affiliation{\Rochester}  \affiliation{\FNAL}
\author{C.L.~McGivern}                    \affiliation{\Pittsburgh}
\author{A.M.~McGowan}                     \affiliation{\Rochester}
\author{B.~Messerly}                      \affiliation{\Pittsburgh}
\author{J.~Miller}                        \affiliation{\USM}
\author{A.~Mislivec}                      \affiliation{\Rochester}
\author{J.G.~Morf\'{i}n}                  \affiliation{\FNAL}
\author{J.~Mousseau}                      \affiliation{\Florida}
\author{T.~Muhlbeier}                     \affiliation{\CBPF}
\author{D.~Naples}                        \affiliation{\Pittsburgh}
\author{J.K.~Nelson}                      \affiliation{\WM}
\author{A.~Norrick}                       \affiliation{\WM}
\author{J.~Osta}                          \affiliation{\FNAL}
\author{V.~Paolone}                       \affiliation{\Pittsburgh}
\author{J.~Park}                          \affiliation{\Rochester}
\author{C.E.~Patrick}                     \affiliation{\Northwestern}
\author{G.N.~Perdue}                      \affiliation{\FNAL}  \affiliation{\Rochester}
\author{L.~Rakotondravohitra}\thanks{\lazaThanks}  \affiliation{\FNAL}
\author{R.D.~Ransome}                     \affiliation{\Rutgers}
\author{H.~Ray}                           \affiliation{\Florida}
\author{L.~Ren}                           \affiliation{\Pittsburgh}
\author{P.A.~Rodrigues}                   \affiliation{\Rochester}
\author{D.~Ruterbories}			  \affiliation{\Rochester}
\author{H.~Schellman}                     \affiliation{\Northwestern}
\author{D.W.~Schmitz}                     \affiliation{\Chicago}  \affiliation{\FNAL}
\author{C.~Simon}                         \affiliation{\UCIrvine}
\author{F.D.~Snider}                      \affiliation{\FNAL}
\author{J.T.~Sobczyk}\thanks{\janThanks}                     \affiliation{\FNAL}
\author{C.J.~Solano~Salinas}              \affiliation{\UNI}
\author{N.~Tagg}                          \affiliation{\Otterbein}
\author{B.G.~Tice}\thanks{\ticeThanks}    \affiliation{\Rutgers}
\author{E.~Valencia}                      \affiliation{\Guanajuato}
\author{J.~Wolcott}                       \affiliation{\Rochester}
\author{M.Wospakrik}                      \affiliation{\Florida}
\author{G.~Zavala}                        \affiliation{\Guanajuato}
\author{D.~Zhang}                         \affiliation{\WM}
\author{B.P.Ziemer}                       \affiliation{\UCIrvine}

%
\collaboration{\minerva  Collaboration}\ \noaffiliation

\date{\today}


\begin{abstract}
A study of charged-current muon neutrino scattering on hydrocarbon
in which the final state includes a muon, at least one proton, and no pions is presented. 
Although this signature has the topology of 
neutrino quasielastic scattering from neutrons, the event sample contains contributions from  
quasielastic and inelastic processes where pions are absorbed in the nucleus.  
The analysis accepts events with muon production angles up to 
70$^{\circ}$ and proton kinetic energies greater than 110~MeV. The cross section, 
when based completely on 
hadronic kinematics, is well-described by a relativistic 
Fermi gas nuclear model including the neutrino event generator modeling 
for inelastic processes and particle transportation through the nucleus. This is in contrast
to the quasielastic cross section based on muon kinematics, which is best described by an extended model
that incorporates multinucleon correlations. This measurement guides the formulation of a complete 
description of neutrino-nucleus interactions that encompasses the hadronic as
well as the leptonic aspects of this process.

\end{abstract}

\ifnum\sizecheck=0  
\maketitle
\fi

Neutrino quasielastic scattering $\nu_{l}N(n) \rightarrow l^{-}p$ on nuclei~$N$ 
is a dominant signal process for neutrino oscillation experiments
and is used to extract information about the axial vector form factor for 
nucleons~\cite{Gran:2006jn,AguilarArevalo:2007ab,Espinal:2007zz,Lyubushkin:2008pe,Nakajima:2011zz}.
The simple final-state topology combined with an assumption that the initial-state 
nucleon is at rest allows for an estimate of the neutrino energy 
from the final-state lepton kinematics alone. This estimate can be altered 
by the fact that the nucleon is bound in a nucleus, which is often modeled 
by assuming that the nucleons are non-interacting within a relativistic 
Fermi gas (RFG). However, measurements based on the final-state lepton on nuclei 
with $A$~$>$~2 over different ranges of four-momentum transfer \QSq are inconsistent 
with the RFG model with $M$$_{A}$~$\sim$~1~GeV in the absolute size of the cross section per nucleon 
and the predicted energy deposition near the vertex of these 
interactions~\cite{Gran:2006jn,AguilarArevalo:2007ab,Nakajima:2011zz,AguilarArevalo:2013hm,Fields:2013zhk,Fiorentini:2013ezn}. 

Many models of nuclear effects attempt to explain these discrepancies by considering  
possible correlations between nucleons. These include short-range correlations 
as observed in electron scattering~\cite{Benhar:2006wy,Arrington:2011xs,Arrington:2012ax}, 
long-range correlations that are modeled with the random phase approximation 
(RPA)~\cite{Graczyk:2003ru,Nieves:2004wx,Valverde:2006yi,Martini:2009uj,Martini:2010ex}, 
and meson exchange currents 
(MEC)~\cite{Martini:2009uj,Martini:2010ex,Nieves:2011pp,Bodek:2011ps,Shen:2012xz,Sobczyk:2012ms,Gran:2013kda}.
Each of these processes changes the event rate and final-state particle
kinematics.

In addition, hadrons produced in neutrino-nucleus interactions can undergo final-state 
interactions (FSI) as they propagate through the nucleus. Consequently, 
a sample including only a lepton and nucleons will invariably  
contain events from inelastic processes. These include $\Delta(1232)$ resonance 
production and decay, where the pion is not observed. Events from both 
inelastic processes with no final-state pions and nucleon-nucleon correlations 
contribute to the measured quasielastic (QE) cross section, but have different kinematics and 
final-state hadron content for the same neutrino energy. Therefore, these events 
can alter the accuracy of any neutrino energy 
estimate~\cite{Leitner:2010kp,Martini:2012fa,Martini:2012uc,Nieves:2012yz,Lalakulich:2012gm,Lalakulich:2012hs,Mosel:2013fxa} 
that neutrino oscillation experiments~\cite{LBNE,Jediny:2014lda} use.  

Additional information is accessible through measurements of the 
hadronic component. Previous 
measurements~\cite{Gran:2006jn,Espinal:2007zz,Lyubushkin:2008pe,Nakajima:2011zz} 
have made a selection on the energy and/or direction of a tracked proton and 
its consistency with the QE hypothesis in order to increase the QE purity of 
the sample. Such a selection will remove events modified by FSI or caused by 
non-QE processes. The presented QE-like analysis specifically retains 
sensitivity to these effects by requiring only that the final-state proton's 
direction and momentum be measured.

Presented is a differential cross-section measurement of QE-like events
that consist of a muon with at least one proton and no pions in the final state. 
By using the kinetic energy of the most energetic (leading) proton, a measurement 
of \QSq is made from the hadronic component alone. This extracted cross section 
is measured with much improved acceptance at large muon scattering angles and 
higher \QSq than that of \minerva's QE measurements that rely on muon 
kinematics~\cite{Fields:2013zhk,Fiorentini:2013ezn}.


The \minerva detector~\cite{Aliaga:2013uqz} is comprised of a fine-grained scintillator 
central tracking region surrounded by electromagnetic and hadronic calorimeters. 
The core consists of tracking planes made of interleaved scintillator 
strips of triangular profile, enabling charged-particle energy depositions 
to be located to within 3~mm. The planes are mounted vertically, nearly perpendicular 
to the neutrino beam axis which is 58~mrad from horizontal.   
Three different plane orientations ($0^{\circ}$ and $\pm 60^{\circ}$ from the vertical) 
permit three-dimensional reconstruction of charged particle trajectories. 
The detector's 3~ns hit-time resolution allows separation of multiple neutrino interactions 
within each 10~$\mu$s spill from the accelerator. The scintillator planes
are supported by exterior hexagonal steel frames with rectangular 
scintillator bars embedded into slots, which serve as the side hadronic calorimeter. 
The magnetized \minos near detector~\cite{Michael:2008bc} located two meters 
downstream of \minerva serves as a muon spectrometer. 

These data were taken in the \numi beamline at Fermilab when its focusing elements 
were configured to produce an intense beam of muon neutrinos (\numu) peaked at 3.5~GeV. 
The run period for these data occurred between March 2010 and April 2012, and 
corresponds to $3.04\times{10}^{20}$ protons on target (POT). The neutrino flux is over 
$95\%$ \numu in the peak, with the remainder consisting of 
\numubar, $\nu_e$, and $\bar\nu_e$, and is predicted using a \geant-based model 
constrained by hadron production data~\cite{Alt:2006fr} 
as described in Ref.~\cite{Eberly:2014mra}.  

Neutrino interactions are simulated using the \genie 2.6.2~\cite{Andreopoulos:2009rq} 
event generator. The propagation of particles in the detector and the corresponding detector response  
are simulated with \geant~\cite{Agostinelli:2002hh}. The calorimetric 
energy scale is tuned using through-going muons to ensure that the photon
statistics and reconstructed energy deposition agree between simulation and data.
Measurements made with a smaller version of the \minerva detector in a low energy 
hadron test beam~\cite{Aliaga:2013uqz} are used to constrain the 
uncertainties associated with the detector response to both protons and charged pions.


For each QE-like candidate, a muon and at least one proton are reconstructed as 
tracks, where the proton track originates from the most upstream position of 
the muon track. This selection includes all muons that exit either the side of the 
central tracking region or the downstream end. Therefore, the selection has
good acceptance for events with muon scattering angles up to 70$^{\circ}$ relative 
to the beam direction. Events are required to occur at least 22~cm from the edge 
of the scintillator and within the central 110~planes of the tracking region, defining a 
fiducial region of 5.57~metric tons. For $53\%$ of the events, the muon track 
is matched to a track in the \minos detector, allowing the charge and momentum to be 
determined. An additional $8\%$ of muons  
entering \minos are not tracked there. As for the muons that exit the 
\minerva outer calorimeter, only a lower-bound on the momentum is obtained. 
A minimum of five distinct energy depositions is required to form a 
proton track, resulting in a 110~MeV kinetic energy threshold. 
The proton tracks must stop in the inner region of \minerva. 

Particle identification (PID) and the reconstructed energy for
protons are determined using a track-based d$E/$d$x$ algorithm.
The algorithm fits the measured d$E/$d$x$ profile of 
a track to predicted profiles for both proton and pion hypotheses, 
and the two fit $\chi^2$ values are used to construct a proton PID consistency score~\cite{Twalton-thesis}. 
This fitting routine successfully identifies protons that
rescatter due to nuclear interactions and provides a kinetic energy
resolution of 5$\%$ for all identified protons. An event is retained 
if all non-muon tracks pass a cut on the proton PID consistency score.  

The remaining cuts are designed to remove inelastic background events with 
an untracked pion. Pions with kinetic energies above 100~MeV 
are likely to interact strongly within the detector materials, produce hadronic 
showers, and consequently are unable to be reconstructed as tracks. These events 
are removed by cutting on energy $E_{extra}$ that is not linked to a track and is located 
outside of a 10~cm sphere centered at the vertex. Excluding this vertex region 
when making this cut reduces sensitivity to mismodeling of low energy 
nucleons~\cite{Fields:2013zhk,Fiorentini:2013ezn}, which may arise from FSI or 
multinucleon effects.  Pions with kinetic energies below 100~MeV are removed 
using an algorithm that 
identifies Michel electrons from the $\pi \to \mu \to e$ decay chain occurring
near the vertex at a delayed time relative to the initial neutrino interaction.
After applying all cuts, the sample contains 40,102 QE-like candidates. 
The simulation predicts that 34.5\% of the events are from 
backgrounds containing at least one final-state pion, where the 
backgrounds are described below.

\begin{figure}[htpb]
\centering
\includegraphics[width=\columnwidth]{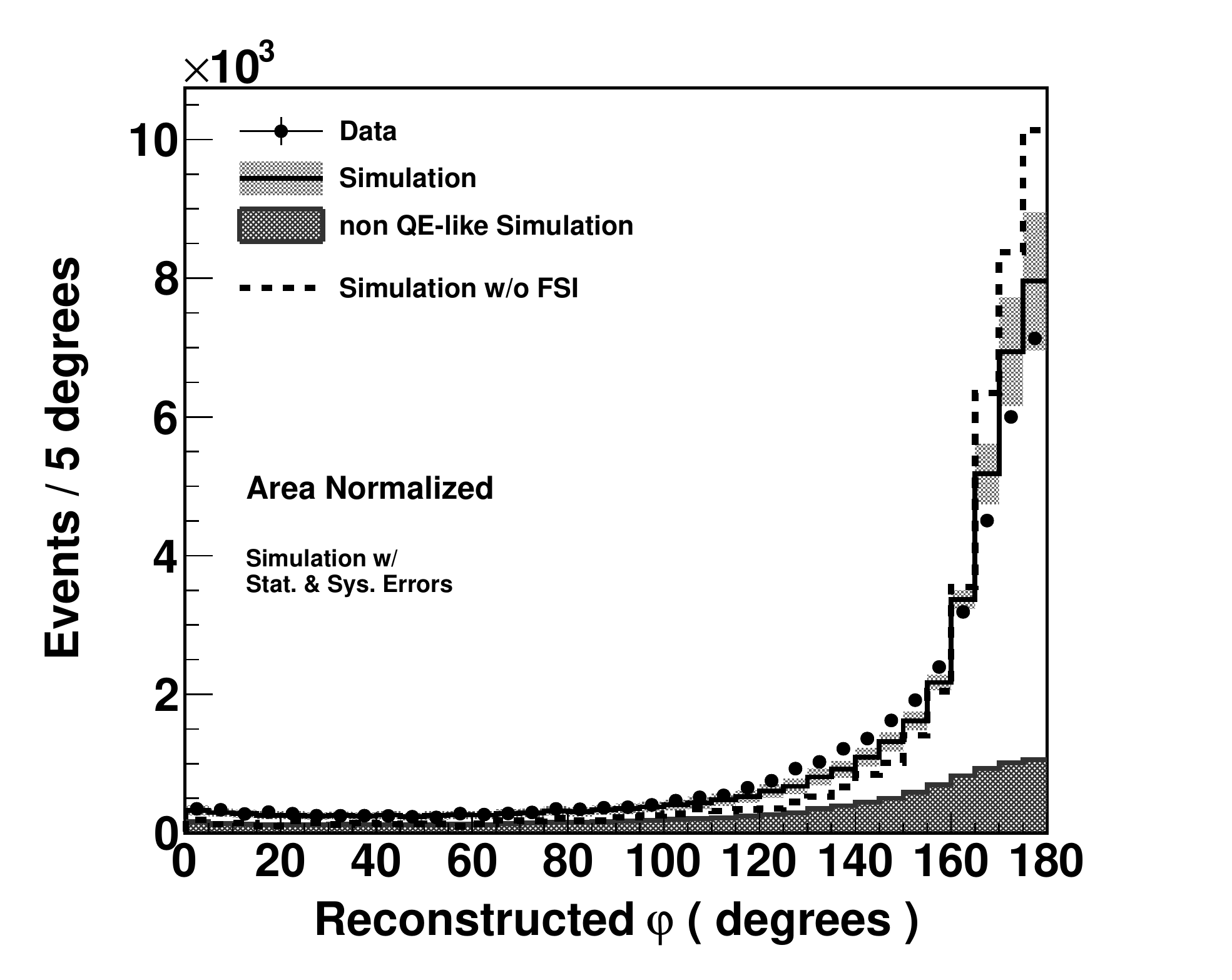}
\caption{Angle between the $\nu$-muon and $\nu$-proton planes for data (black points) and two
predictions from \genie, where the solid line prediction includes FSI
and the dashed line prediction does not. The total predictions have been normalized to the
data, and the non-QE-like predictions have been normalized to sidebands in the data.}
\label{fig:muon_proton_compare_angle}
\end{figure}

Measurement of the proton angle and momentum provides several variables 
that are sensitive to FSI. One variable is 
the angle~$\varphi$ between the $\nu$-muon and $\nu$-proton planes, and 
is shown in Fig.~\ref{fig:muon_proton_compare_angle} for both the data and 
two simulations: one with FSI and one without FSI. 
For both simulations, the non-QE-like 
background is tuned using a data-based procedure described below.
For QE scattering off a 
free neutron at rest $\varphi=180^\circ$. The detector resolution on 
$\varphi$ is 3.8~degrees, so the width shown on the distributions in 
Fig.~\ref{fig:muon_proton_compare_angle} is due to Fermi 
motion, inelastic scattering, and FSI effects. The comparison shows that \genie with 
FSI describes the data better than \genie without FSI. The remaining discrepancy 
suggests additional FSI or cross section effects not present in the \genie simulation.  

\begin{figure}[htpb]
\centering
\includegraphics[width=\columnwidth]{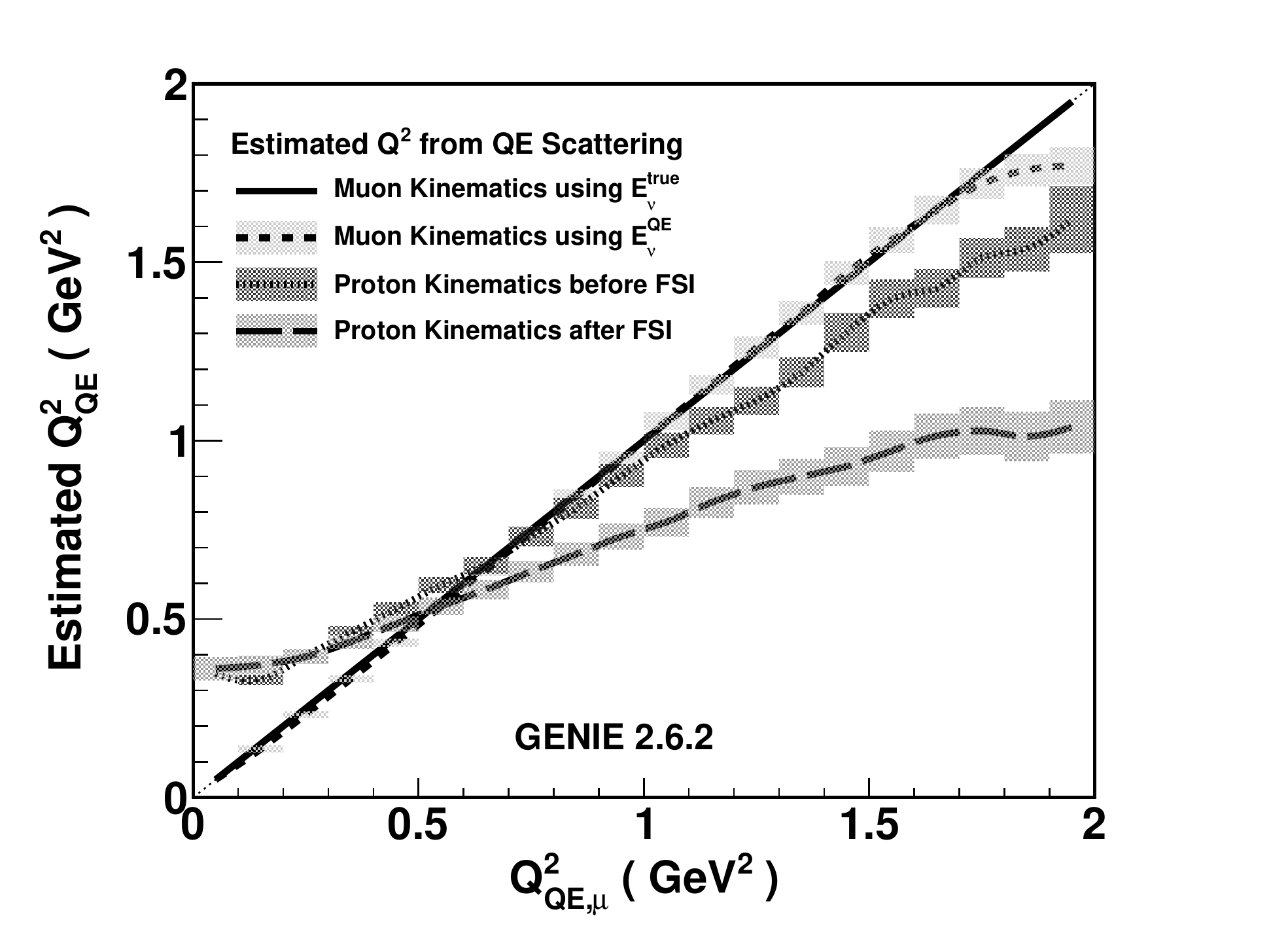}
\caption{Comparisons between several \QSq estimates as function of the \QSq estimated
using muon kinematics for the QE-like signal events that pass the reconstruction
and analysis cuts, as described in the text. The error bands include 
statistical and \genie systematic uncertainties.}
\label{fig:q2_correlation}
\end{figure}

The differential cross section \dsdq is measured using the leading proton 
and the assumption of QE scattering from a neutron at rest. 
Under this assumption, \QSq is given by 
\begin{equation*}
\QSqproton=(M_{n}-\epsilon_B)^2-M^2_{p}+2(M_{n}-\epsilon_{B})(T_{p}+M_{p}-M_{n}+\epsilon_{B}),
\end{equation*}
where $T_{p}$ is the kinetic energy of the proton, $M_{n,p}$ is the nucleon mass, 
and $\epsilon_B$ is the effective binding energy of +34 MeV~\cite{Moniz:1971mt}. 
This estimation of \QSqproton depends only on the $T_{p}$ of the leading proton.
This approximation of 
\QSqproton deviates from the \QSq estimated using only the muon. 
For the QE-like signal events that pass the analysis cuts, Fig.~\ref{fig:q2_correlation} 
shows \genie's  average values of various estimates of \QSq using truth information
as a function of \QSq as defined by the muon kinematics, namely: 
\begin{equation*}
\QSqmuon=-m^{2}_{\mu}+2E_{\nu}(E_{\mu}-\sqrt{(E^{2}_{\mu}-m^{2}_{\mu})}\cos\theta_{\mu}),
\end{equation*}
where $E_{\mu}$, $\theta_{\mu}$, and $m_{\mu}$ are the true energy, 
true scattering angle, and mass of the muon and $E_{\nu}$ is the true 
energy of the neutrino. The solid and short-dashed curves show \QSqmuon 
from the muon, and the discrepancy at \QSqmuon~$>$~1.7~GeV$^{2}$ is from
differences in the way the neutrino energy is estimated. The solid curve uses the 
neutrino's true energy, and short-dashed curve uses the QE hypothesis 
to estimate the neutrino energy using the muon's true energy and angle, which is given by 
\begin{equation*}
E_{QE,\nu}=\frac{2(M_{n}-\epsilon_{B})E_{\mu}-[(M_{n}-\epsilon_{B})^{2}+m^{2}_{\mu}-M^{2}_{p}]}{2(M_{n}-\epsilon_{B}-E_{\mu}+\sqrt{(E^{2}_{\mu}-m^{2}_{\mu})}\cos\theta_{\mu})}.
\end{equation*}
At higher \QSqmuon, the QE hypothesis inaccurately describes the inelastic 
component of the QE-like signal. 

The dotted and long-dashed curves show
\QSqproton from the proton, and the effects of FSI contribute to the discrepancy 
between the curves. The tracking threshold prevents the reconstruction of events 
with a leading proton having $T_{p}$~$<$~110~MeV, thereby resulting in a \QSqproton 
limit roughly 0.2~GeV$^{2}$ and poor acceptance for \QSqmuon~$<$~0.2~GeV$^{2}$.
Based on the Bodek-Ritchie~\cite{Bodek:1980ar,Bodek:1981wr} prescription,
\genie models the momentum distribution of initial-state nucleons by including a 
high-momentum tail extending beyond the Fermi momentum. Consequently at low \QSqmuon,
the analysis preferentially selects events where the initial-state 
nucleon momentum is greater than the Fermi momentum. This is a feature of the 
proton-based curves in Fig.~\ref{fig:q2_correlation}. 
The differences between the muon-based and proton-based estimates come from 
Fermi motion for the dotted curve and Fermi motion with FSI for the long-dashed curve, 
where such nuclear effects distort the shape of the \QSqproton distribution.

\begin{figure}[tp]
\centering
\includegraphics[width=\columnwidth]{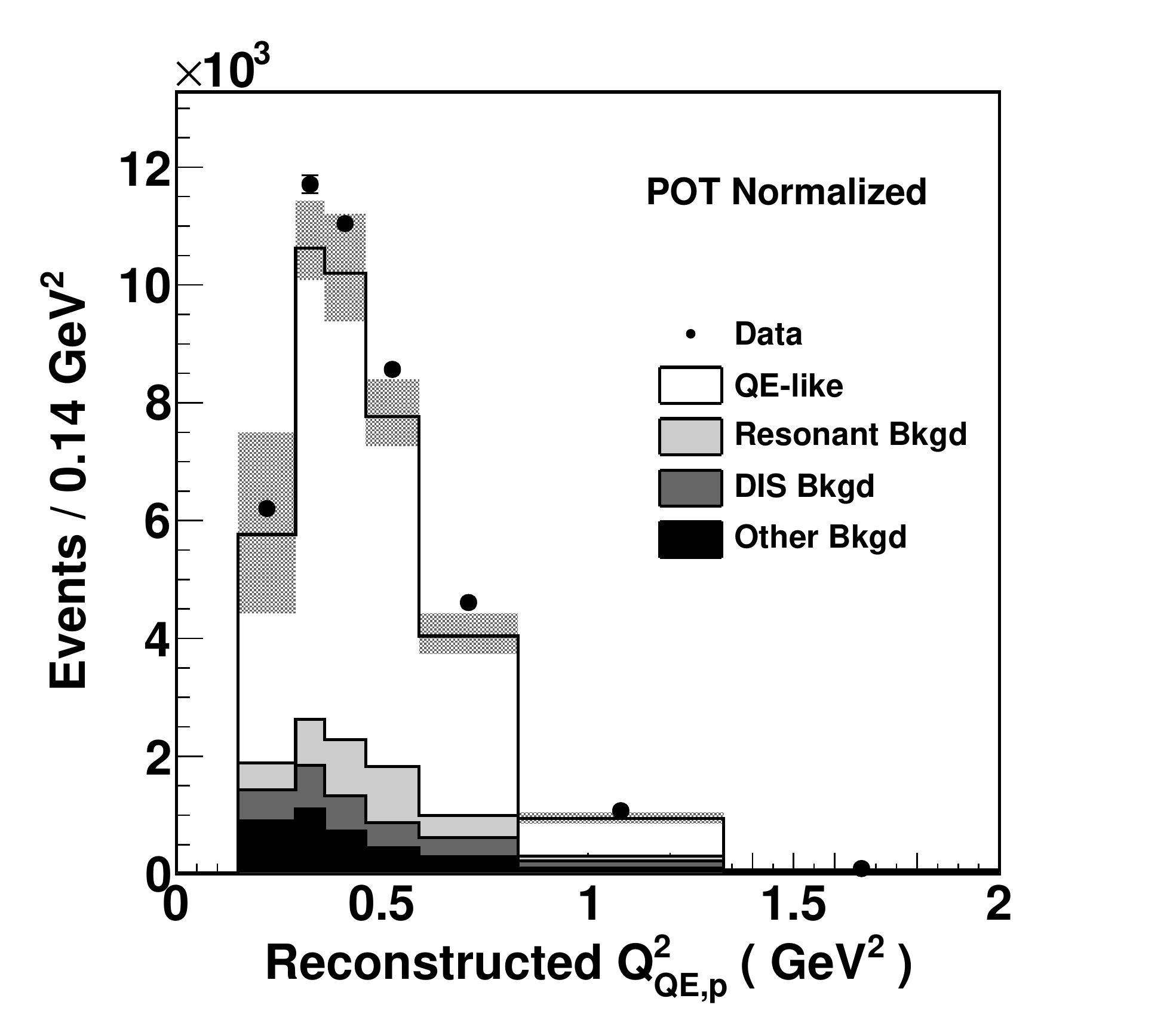}
\caption{Distribution of $Q^2$ of the QE-like events determined by the leading proton 
track reconstruction in data and simulation, where the background estimates are 
tuned to sideband samples of the data. }
\label{fig:ccqelike_reco_q2_pot}
\end{figure}

The reconstructed \QSqproton distribution is shown in Fig.~\ref{fig:ccqelike_reco_q2_pot}. 
The data are compared to \genie predictions of the QE-like signal and 
backgrounds, where the background estimates have been tuned using the data.
The largest background comes from baryon resonance 
production, predominantly $\Delta(1232)$. A smaller background originates
from non-resonant inelastic pion production, which is referred to as
deep-inelastic scattering (DIS) in \genie.
In the tuning procedure, the non-QE-like backgrounds are categorized 
as one of two processes: ``resonant'' or ``DIS plus other'' (hereafter denoted DIS+), 
where the ``other'' includes $\overline{\nu}_{\mu}$ interactions
and a smaller neutral current component. Four distinct sideband regions of the 
$E_{extra}$ distribution are used to extract normalization constants for each process. 
The sidebands further from the signal region are predicted to contain a larger estimated fraction 
of DIS+ relative to the resonant fraction and are more consistent with the data. 
By separating the simulated backgrounds into two processes, 
the backgrounds in each \QSqproton bin are determined from a 
linear fit that simultaneously matches the simulated background to data 
in all sideband regions. The tuning results indicate that 
the baryon resonance production background should be reduced by roughly 50$\%$. 
The DIS+ background prediction remains nearly unchanged for \QSqproton~$>$~0.5~GeV$^{2}$ 
and increases by 20$-$60$\%$ in \QSqproton regions between 0.15 to 0.5~GeV$^{2}$.

After subtracting the data-tuned backgrounds, the yield is corrected for detector 
smearing of the leading proton energy via a Bayesian unfolding procedure using four 
iterations~\cite{D'Agostini:1994zf}. The simulation is used to correct for geometric 
acceptance and efficiency for the unfolded distribution. To obtain the flux-averaged 
differential cross section, the yields are divided by the number of nucleons in the 
fiducial volume (3.294~$\times$~$10^{30}$) and the integrated \numu  
flux below 100~GeV (3.286~$\times$~$10^{-8}\rm /cm^{2}/POT$).


The systematic uncertainties on \xsecproton arise from  
imperfect knowledge of the (I) neutrino beam flux, (II) neutrino interactions, 
(III) final state interactions, (IV) detector energy response, 
(V) hadron inelastic cross sections, and (VI) other sources, and 
are listed in Table~\ref{tab:systematic}.  Most uncertainties are
evaluated by randomly varying the associated parameters 
in the simulation within uncertainties and re-extracting 
\xsecproton. Each variation is normalized to the measured
\xsecproton to extract the uncertainty in the shape. Consequently, in regions
where the shape of the \xsecproton changes dramatically, the uncertainty on
the shape may exceed that of the absolute uncertainty. 

The uncertainty on the beam flux affects the normalization of \xsecproton 
and is correlated across \QSqproton bins. The uncertainties on the 
neutrino interaction and FSI models affect \xsecproton through the efficiency 
correction and are dominated by uncertainties on the resonance production axial mass 
parameter, pion absorption, and pion inelastic scattering. The uncertainties
associated with hadron propagation through \minerva 
are evaluated by shifting the pion and proton total inelastic cross sections 
by 10$\%$, an uncertainty derived from external hadron production 
data~\cite{Lee:2002eq,Ashery:1981tq,Allardyce:1973ce,Saunders:1996ic}. 
This uncertainty affects proton tracking and PID efficiencies and acceptance 
for \QSqproton~$>$~0.8~GeV$^{2}$. 
The systematic uncertainty from the detector energy response is relatively small and 
is dominated by uncertainties on the reconstruction of the proton and $E_{extra}$.

\begin{table}[!htbp]
\centering
\caption{Fractional systematic uncertainties (in units of percent) on \xsecproton for each 
\QSqproton bin, with contributions from (I) neutrino beam flux, (II) neutrino interaction models, 
(III) final-state interaction models, (IV) detector energy response,
(V) hadronic inelastic cross section model, and (VI) other sources. 
The absolute uncertainties are followed by the shape uncertainties in parentheses.}
\scalebox{0.8}
{
\begin{tabular}{ c  c  c  c  c  c  c  c }
\hline
\hline
\QSqproton (GeV$^{2}$) & I         & II        & III       & IV        & V         & VI        & Total  \\ \hline
0.15 - 0.29           & 7.2(0.6)  & 5.3(6.6)  & 4.3(11)   & 3.0(3.9)  & 2.0(3.4)  & 2.9(1.6)  & 11(14) \\
0.29 - 0.36           & 7.6(0.2)  & 5.8(3.1)  & 7.3(1.5)  & 1.3(1.7)  & 3.2(4.7)  & 2.6(1.4)  & 13(6.3) \\
0.36 - 0.46           & 7.5(0.4)  & 7.5(2.0)  & 11(3.2)   & 2.0(1.1)  & 3.3(3.3)  & 1.1(0.3)  & 16(5.1) \\
0.46 - 0.59           & 7.7(0.2)  & 8.8(2.4)  & 13(4.3)   & 2.9(1.1)  & 2.0(0.8)  & 1.0(0.5)  & 17(5.1) \\
0.59 - 0.83           & 8.0(0.3)  & 9.6(3.1)  & 13(4.3)   & 4.1(2.1)  & 1.6(0.9)  & 1.0(0.6)  & 18(5.9) \\
0.83 - 1.33           & 8.2(0.6)  & 10(3.4)   & 12(3.3)   & 7.5(5.6)  & 9.6(8)    & 1.4(2.1)  & 21(11) \\
1.33 - 2.00           & 8.2(0.7)  & 11(4.3)   & 11(2.4)   & 7.8(7.5)  & 3.8(3.2)  & 1.9(1.2)  & 19(9.6) \\ 
\hline
\hline
\end{tabular}
}
\label{tab:systematic}
\end{table}

\begin{figure}[htpb]
\centering
\includegraphics[width=\columnwidth]{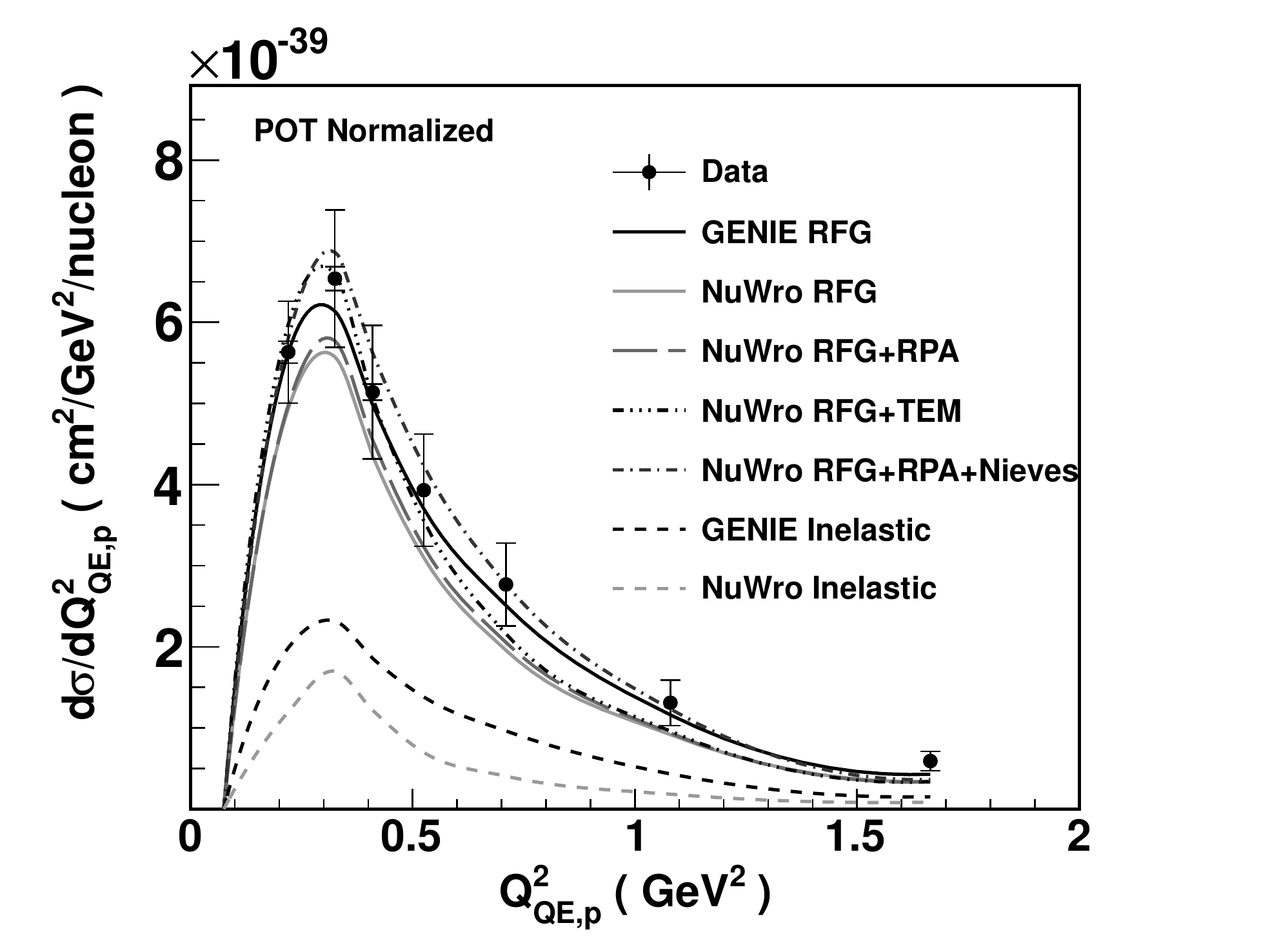}
\includegraphics[width=\columnwidth]{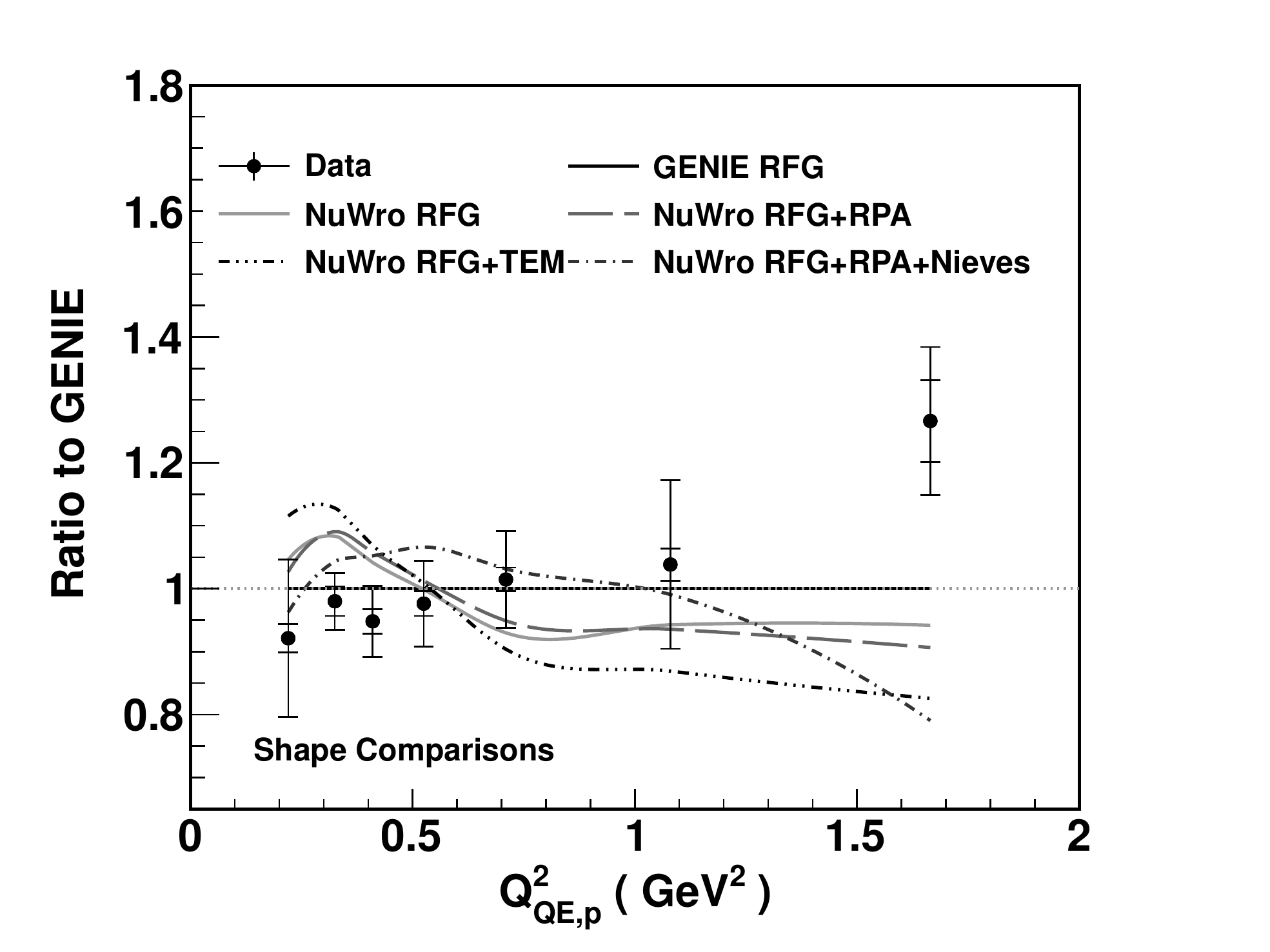}
\caption{(Top) QE-like cross section versus \QSqproton compared to several different 
predictions, along with the \genie (dark dashed line) and \nuwro (light dashed line) 
predictions of the inelastic contribution to the QE-like prediction.
(Bottom) Ratio between the data and predictions to the \genie RFG prediction including the 
inelastic component, where all models are normalized to the data.  
The inner (outer) error bars correspond to the 
statistical (total) uncertainties.}
\label{fig:cross_section_model_overlay_POT}
\end{figure}

The QE-like differential cross section~\cite{supplemental} as a function of 
\QSqproton is shown in Fig.~\ref{fig:cross_section_model_overlay_POT} (top), 
along with predictions from the \genie and \nuwro~\cite{Nowak:2006xv,Golan:2013jtj} 
generators using a RFG model. Several extensions to the 
\nuwro QE RFG prediction are shown, inspired by measurements 
based on lepton kinematics. Each prediction represents the sum over all 
reactions with at least one $>$110~MeV proton and no pions in 
the final state. The inelastic contributions to this QE-like cross section  
from the \genie (dark dashed) and \nuwro (light dashed) predictions 
are shown, and differ in both rate and shape. For both generators, 
the inelastic component is dominated by $\Delta(1232)$ production and decay, 
where the pion is absorbed by the residual nucleus. 

The shape of \xsecproton can be compared between prediction
and data with reduced systematic uncertainty by normalizing each
prediction to the data. Figure~\ref{fig:cross_section_model_overlay_POT} (bottom) 
shows the ratio between the data and normalized prediction to the \genie
RFG prediction. For both the absolute and the shape comparisons, the $\chi^2$ 
between the data and each prediction, including correlations between bins, 
is given in Table~\ref{tab:model_compare}. The highest \QSqproton data point
contributes mostly to the total $\chi^2$ for the \genie RFG and \nuwro RFG 
with RPA+Nieves. The remaining predictions get most of their
total $\chi^2$ from the middle data points, which have higher
and positive covariance values, but these predictions visibly have an opposite 
trend to the data.

The rate and shape of the data are best described by the \genie RFG model 
with the inelastic component, followed 
by the \nuwro RFG model with its very different prediction of the 
inelastic process. The remaining \nuwro predictions 
become more discrepant with the shape of the data, as various implementations 
of nuclear effects are incorporated. 

\genie and \nuwro calculate QE scattering using the independent 
nucleon impulse approximation, 
the BBBA2005 parameterization~\cite{Bradford:2006yz} 
of the vector form factors, and an axial mass of $0.99~\rm GeV/c^2$. 
As described above, \genie has in addition an approximation of short-range 
correlations included 
as prescribed by Bodek-Ritchie~\cite{Bodek:1980ar,Bodek:1981wr}.

\begin{table}[!htbp]
\centering
\caption{Calculated $\chi^{2}$ between the data and various models with M$_{A}$ = 0.99~GeV/c$^{2}$. 
The number of degrees of freedom is 7 (6) for the rate (shape).\label{tab:model_compare}}
\scalebox{0.8}
{
\begin{tabular}{ l c c }
Model & Rate $\chi^{2}$   & Shape $\chi^{2}$ \\ \hline
\genie RFG                &   8.5  &  10.8 \\
\nuwro RFG                &  12.2  &  19.9 \\
\nuwro RFG + RPA          &  13.5  &  21.7 \\
\nuwro RFG + RPA + Nieves &  25.9  &  28.5 \\
\nuwro RFG + TEM          &  27.6  &  34.5 \\
\end{tabular}
}
\end{table}

The difference between \genie and \nuwro RFG predictions arises primarily in
the difference between the two simulations' treatments of inelastic scattering.
In \genie, resonance production is defined using the formalism of 
Rein-Sehgal~\cite{Rein:1980wg} while in \nuwro, 
it is defined as interactions with invariant hadronic mass $W <$~1.6~GeV, 
and where contributions come from $\Delta(1232)$ 
excitations. \genie and \nuwro handle interactions with $W >$~1.6~GeV similarly 
and transport hadrons through the nucleus using different implementations of 
an intranuclear cascade model. 

RPA calculations predict a suppression of the cross section at very low \QSq 
from long-range correlations, and enhancement at moderate \QSq 
due to short-range correlations. For the \nuwro RPA 
calculation~\cite{Graczyk:2003ru}, the suppression happens below \minerva's proton 
kinetic energy threshold, and its curve is nearly identical to \nuwro RFG 
prediction as shown in Fig.~\ref{fig:cross_section_model_overlay_POT}.
MEC between nucleons enhance the cross section and populate the transition 
region between the QE and $\Delta$ peaks. MEC are part of 
both the microscopic model of Nieves~\cite{Nieves:2004wx,Gran:2013kda} and 
the Transverse Enhancement Model (TEM) which is empirically extracted from 
electron scattering data~\cite{Bodek:2011ps}. TEM is based on QE 
lepton kinematics so that the energy transfer follows from reweighting QE events 
rather than filling the transition region between the QE and $\Delta$ peaks, while the Nieves 
model gives systematically higher energy transfers which translate to an enhancement 
at higher proton energies. The proton kinematics are not explicitly calculated  
in either the TEM or the Nieves model, and in NuWro are assigned as described in 
Refs.~\cite{Sobczyk:2012ms,Golan:2013jtj}.

The agreement between the presented QE-like data and the \genie prediction
is in stark contrast to that of the \minerva QE 
measurements~\cite{Fields:2013zhk,Fiorentini:2013ezn}, where \QSq is based on 
muon kinematics and the backgrounds from all inelastic events 
are subtracted. To check the consistency between 
\minerva muon-based QE and proton-based QE-like measurements, the subsample of 
events with muons that are tracked in \minos is used to measure the pure QE 
differential cross section~\cite{supplemental} as a function of \QSq estimated 
from the muon kinematics as given in Ref~\cite{Fiorentini:2013ezn}. 
These results are consistent with the reported QE measurement~\cite{Fiorentini:2013ezn} 
while using a factor of three more protons on target but lower acceptance because 
of the proton track requirement.

The inconsistency of the models of the hadronic and leptonic aspects of the
QE-like sample may be resolved by modifying the $\Delta(1232)$ production cross section and 
nuclear absorption models. Supporting evidence comes from the results of 
the background tuning described above and \minerva's inclusive pion production
measurement~\cite{Eberly:2014mra},which finds $\Delta$-dominated single pion production 
to be nearly 30\% less than the \genie prediction. Refinements to models of multinucleon 
effects, beyond those implemented in these versions of \genie and \nuwro, may 
also resolve the discrepancies seen here.  

This proton-based \xsecproton measurement provides a new way to evaluate the 
modeling of all contributions to the QE-like cross section, and finds that the 
models which best describe the proton kinematics of this interaction differ
from those that best describe the muon kinematics. The models used by 
neutrino oscillation experiments must ultimately reproduce the hadronic as well as 
leptonic kinematics since both affect neutrino energy reconstruction.

\ifnum\sizecheck=1
  \newpage
  {\Large Content after here does not count against size of PRL}
  \newpage
\fi
\begin{acknowledgments}

This work was supported by the Fermi National Accelerator Laboratory
under U.S. Department of Energy Contract
No.\@ DE-AC02-07CH11359 which included the \minerva construction project.
Construction support also
was granted by the United States National Science Foundation under
Grant No. PHY-0619727 and by the University of Rochester. Support for
participating scientists was provided by NSF and DOE (USA) by CAPES
and CNPq (Brazil), by CoNaCyT (Mexico), by CONICYT (Chile), by
CONCYTEC, DGI-PUCP and IDI/IGI-UNI (Peru), by Latin American Center for
Physics (CLAF), by the Swiss National Science Foundation, and by RAS and the Russian Ministry of Education and Science (Russia).  We
thank the MINOS Collaboration for use of its
near detector data. Finally, we thank the staff of
Fermilab for support of the beam line and detector.

\end{acknowledgments}

\bibliographystyle{apsrev4-1}
\bibliography{ProtonCCQElike.bib}

\ifnum\PRLsupp=0
  \clearpage
  \newcommand{\qsq}{\ensuremath{Q^2_{QE}}\xspace}
\renewcommand{\textfraction}{0.05}
\renewcommand{\topfraction}{0.95}
\renewcommand{\bottomfraction}{0.95}
\renewcommand{\floatpagefraction}{0.95}
\renewcommand{\dblfloatpagefraction}{0.95}
\renewcommand{\dbltopfraction}{0.95}
\setcounter{totalnumber}{5}
\setcounter{bottomnumber}{3}
\setcounter{topnumber}{3}
\setcounter{dbltopnumber}{3}

\ifdefined\PRLsupp
  \ifnum\PRLsupp=0
    \newcommand{\SuppName}{appendix}
  \else
    \newcommand{\SuppName}{supplemental material}
  \fi
\else
  \newcommand{\SuppName}{supplemental material}
\fi

\section{\SuppName}

\begingroup
\squeezetable
\begin{table}[h]
\begin{tabular}{c|c c c c c c c c c c c c c}
\hline
\hline 
$E_\nu$ in Bin & 
$0 - 0.5$ &
$0.5 - 1$ &
$1 - 1.5$ &
$1.5 - 2$ &
$2 - 2.5$ &
$2.5 - 3$ &
$3 - 3.5$ &
$3.5 - 4$ \\
$\nu_\mu$ Flux
(neutrinos/cm$^2$/POT ($\times 10^{-8}$)) &
$0.039	$&
$0.107	$&
$0.251	$&
$0.360  $&
$0.475	$&
$0.579	$&
$0.594	$&
$0.462	$ \\
\hline
$E_\nu$ in Bin & 
$4 - 4.5$ &
$4.5 - 5$ &
$5 - 5.5$ &
$5.5 - 6$ &
$6 - 6.5$ &
$6.5 - 7$ &
$7 - 7.5$ &
$7.5 - 8$ \\
$\nu_\mu$ Flux
(neutrinos/cm$^2$/POT ($\times 10^{-8}$)) &
$0.275	$&
$0.149	$&
$0.089	$&
$0.062  $&
$0.048	$&
$0.041	$&
$0.035	$&
$0.031	$ \\
\hline
$E_\nu$ in Bin & 
$8 - 8.5$ &
$8.5 - 9$ &
$9 - 9.5$ &
$9.5 - 10$ &
$10 - 11$ &
$11 - 12$ &
$12 - 13$ &
$13 - 14$ &
$14 - 15$ \\
$\nu_\mu$ Flux
(neutrinos/cm$^2$/POT ($\times 10^{-8}$)) &
$0.028  $&
$0.025	$&
$0.023	$&
$0.020	$&
$0.033	$&
$0.028	$&
$0.023	$&
$0.020	$&
$0.017	$ \\ \hline
$E_\nu$ in Bin & 
$15 - 20$ &
$20 - 25$ &
$25 - 30$ &
$30 - 35$ &
$35 - 40$ &
$40 - 45$ &
$45 - 50$ &
$50 - 55$ &
$55 - 60$ \\
$\nu_\mu$ Flux
(neutrinos/cm$^2$/POT ($\times 10^{-9}$)) &
$0.480  $&
$0.204	$&
$0.097	$&
$0.066	$&
$0.052	$&
$0.039	$&
$0.027	$&
$0.015	$&
$0.006	$ \\ \hline
$E_\nu$ in Bin & 
$60 - 65$ &
$65 - 70$ &
$70 - 75$ &
$75 - 80$ &
$80 - 85$ &
$85 - 90$ &
$90 - 95$ &
$95 - 100$&
\\
$\nu_\mu$ Flux
(neutrinos/cm$^2$/POT ($\times 10^{-11}$)) &
$0.310  $&
$0.172	$&
$0.108	$&
$0.061	$&
$0.030	$&
$0.015  $&
$0.008	$&
$0.004	$&
\\ 
\hline
\hline
\end{tabular}
\caption{The calculated muon neutrino flux per proton on target (POT) for the data included in this analysis.}
\end{table}
\endgroup

\begingroup
\squeezetable
\begin{table}[ht]
\begin{tabular}{c | c c c c c c c }
\hline
\QSqproton Bins (GeV$^{2}$) & 
0.15 - 0.29 & 
0.29 - 0.36 & 
0.36 - 0.46 & 
0.46 - 0.59 & 
0.59 - 0.83 & 
0.83 - 1.33 & 
1.33 - 2.0 \\ 
\hline
Cross-section in bin &               
0.563 &
0.654 &
0.514 &
0.393 & 
0.277 & 
0.131 & 
0.059 \\
(10$^{-38}$cm$^{2}$/GeV$^{2}$/nucleon) &
$\pm$0.013$\pm$0.061 &
$\pm$0.015$\pm$0.083 & 
$\pm$0.010$\pm$0.082 & 
$\pm$0.007$\pm$0.069 &
$\pm$0.005$\pm$0.051 &
$\pm$0.003$\pm$0.028 &
$\pm$0.003$\pm$0.011 \\
\hline
\QSqproton Bins (GeV$^{2}$) & 
& 
& 
& 
& 
& 
& 
\\ 
\hline
0.15 - 0.29 & 1 & 0.48 & 0.24 & 0.19 & 0.19 & 0.14 & 0.26 \\ 
0.29 - 0.36 &   & 1    & 0.86 & 0.86 & 0.85 & 0.65 & 0.69 \\ 
0.36 - 0.46 &   &      & 1    & 0.94 & 0.93 & 0.78 & 0.82 \\ 
0.46 - 0.59 &   &      &      & 1    & 0.98 & 0.88 & 0.80 \\ 
0.59 - 0.83 &   &      &      &      & 1    & 0.90 & 0.80 \\ 
0.83 - 1.33 &   &      &      &      &      & 1    & 0.72 \\ 
1.33 - 2.0  &   &      &      &      &      &      & 1    \\ 
\hline
0.15 - 0.29 & 1 & 0.52 & -0.36 & -0.83 & -0.86 & -0.63 & -0.24 \\ 
0.29 - 0.36 &   & 1    &  0.10 & -0.22 & -0.33 & -0.75 & -0.30 \\ 
0.36 - 0.46 &   &      &  1    &  0.42 &  0.26 & -0.23 &  0.17 \\ 
0.46 - 0.59 &   &      &       &  1    &  0.82 &  0.41 &  0.02 \\ 
0.59 - 0.83 &   &      &       &       &  1    &  0.52 &  0.03 \\ 
0.83 - 1.33 &   &      &       &       &       &  1    & -0.05 \\ 
1.33 - 2.0  &   &      &       &       &       &       &  1    \\
\hline
\end{tabular}
\caption{The measured \xsecproton (top) in \QSqproton for QE-like events having $p_{p} >$~110~MeV
of kinetic energy, 
where the first uncertainty is statistical and the second is systematic. In addition, the
full (middle) correlation matrix for the uncertainities on the proton-based cross section measurement.
(Bottom) Corresponding shape correlation matrix.} 
\end{table}
\endgroup

\begingroup
\squeezetable
\begin{table}[ht]
\begin{tabular}{ c c }
\QSqmuon   &  Cross-section \\ 
(GeV$^{2}$) & (10$^{-38}$cm$^{2}$/GeV$^{2}$/neutron) \\ 
\hline
0.2 - 0.4 & 1.164 $\pm$0.035 $\pm$0.193   \\
0.4 - 0.8 & 0.534 $\pm$0.013 $\pm$0.071   \\
0.8 - 1.2 & 0.222 $\pm$0.011 $\pm$0.031   \\
1.2 - 2.0 & 0.110 $\pm$0.011 $\pm$0.015   \\
\end{tabular}
\caption{The cross section as a function of \QSqmuon as measured from the 
muon kinematics for events with an identified proton and a muon that has 
been analyzed by the MINOS detector. The first uncertainty is statistical and 
the second is systematic.\label{tab:minos_xsec}}
\end{table}
\endgroup

\fi

\end{document}